# Toward Agentic Environments: GenAI and the Convergence of AI, Sustainability, and Human-Centric Spaces

**Przemek Pospieszny**
MindLab, EPAM Systems, Newtown, USA
przemek_pospieszny@epam.com

**Dominika P. Brodowicz**
Real Estate and Innovative City Department, Warsaw School of Economics, Poland
dominika.brodowicz@sgh.waw.pl

**Abstract**

In the past few years evolution of artificial intelligence (AI), particularly generative AI (GenAI) and large language models (LLMs), made human-computer interactions more frequent, easier and faster than ever before. This brings numerous benefits in terms of enhancing efficiency, accessibility, and convenience in various sectors from banking to health. AI tools and solutions applied in computers and communication devices support decision-making processes and managing operations of users on the individual level as well as organisational including resource allocation, workflow automation, and real-time data analysis. However, the current use of AI carries a substantial environmental footprint due to its reliance on high-computational cloud resources. In such a context, this paper introduces the concept of agentic environments, a sustainability-oriented AI framework that goes beyond reactive systems by leveraging GenAI, multi-agent systems, and edge computing to minimize negative impact of technology. These types of environments can contribute to the optimisation of resource use, enhanced quality of life, and prioritization of sustainability, while at the same time safeguarding user privacy through decentralized, edge-driven AI solutions. Based on both secondary and primary data gathered during a focus group and semi-structured interviews with AI professional from leading technology companies, authors provide a conceptual framework of agentic environments and discuss it in the context of three lenses including personal sphere, professional and commercial use and urban operations. The findings include the potential of agentic environments to foster sustainable ecosystems, mainly due to optimising of resources usage and securing privacy of data. The study outlines recommendations for implementing edge-driven deployment models to reduce dependency on currently widely applied high-energy cloud solutions.

## 1. Introduction

Over the past two decades, human-computer interactions in various walks of life have been revolutionised through use of smart devices, sensors and most recently by artificial intelligence (AI). This progress is visible through the common application of intelligent environments and AI applications in our homes, leisure, travel, work and education, with an increasing emphasis on local resource efficiency and environmental consciousness [1]. Firstly smart environments contributed to sustainability by automating energy management and optimizing tasks using real-time data analytics to reduce waste and improve consumption. Going further, ambient intelligence offered context-aware services that adapt to user behaviour, indirectly reducing resource use by aligning with human needs [4]. Intelligent environments, the most advanced iteration, incorporate cognitive modelling and AI collaboration to balance user needs with the ecological impact of these actions and ultimately it offers adaptive solutions for sustainable living [5,6].

Recent breakthroughs in generative AI and the increasing application of multimodal large language models (LLMs) as digital assistants have highlighted a significant issue. GenAI represents a paradigm shift in AI, unlocking unprecedented potential for creating interconnected, intelligent environments through compound AI and multi-agent systems capable of seamless interaction with various devices and sensors. These systems go beyond language-based tasks and the integration of web or enterprise systems, allowing for user engagement with the physical world and enabling dynamic, real-world interactions through Physical AI. However, the current use of GenAI carries a substantial sustainability footprint due to its reliance on high-computational cloud resources and the absence of sustainability-aware frameworks that would support its deployment in human-cantered environments. It is estimated that a single ChatGPT prompt consumes energy equivalent to powering a 60-watt incandescent light bulb for approximately 17 minutes. Even a basic prompt requires 10 kilojoules of energy, roughly ten times that of a typical Google search [7]. Training a basic LLM is estimated to emit 552 metric tons of $CO_2$, which is comparable to the annual emissions of 123 gasoline-powered passenger vehicles [8]. Additionally, both the training and inference of AI models contribute to a high demand for water, with projected usage potentially reaching half of the United Kingdom's water consumption by 2027 [9]. While Google anticipates a 48% increase in emissions over the next years, recent investigations have found that inaccuracies in chatbot responses can increase energy use by prompting multiple clarifying queries, further compounding the environmental toll of AI deployments [10]. In parallel, a massive data breach affecting over 16 billion accounts from major firms like Apple, Facebook, and Google has raised urgent questions about AI-related vulnerabilities and the critical need for robust digital security [11].

In such a context, this paper introduces the concept of agentic environments, a sustainability-oriented AI framework that goes beyond reactive systems by integrating personal, professional, and urban domains with GenAI, multi-agent systems, and edge computing to minimize environmental footprint of technology. These environments have a potential to optimise use of resources, enhance quality of life, and prioritise sustainability through decentralised, edge-driven AI solutions. Presented research argues for sustainable AI frameworks in human-centric environments and explores the technological foundations, architecture, and applications of agentic environments as a roadmap for creating intelligent spaces. The study demonstrates how this integrated approach can transform daily operations, such as workplace, commuting, and leisure, while addressing key concerns like privacy and security of AI deployment by examining personal, professional.

## 2. Methodology

The methodological approach applied in this study is of an explorative nature. In general, an explorative study is a type of research focused on insights into a relatively new and unexplored area of interest, and identifying emerging patterns [12], particularly when there is limited existing literature or when new variables are being considered [13]. In this study, the focus is not only on generative AI but also on the concept of agentic environments. The explorative approach offers a flexible, open-ended framework that allows researchers to examine phenomena without predefined hypotheses or expectations [14]. This study was based on both primary and secondary data, including desktop research focused on the analysis of evolution of technological solutions and was structured into three interconnected phases.

The first stage of study includes a comprehensive review of academic literature based on IEEE Xplore, Scops and Web of Science search engines, white papers, industry reports, and regulatory frameworks related to smart environments and AI implementation. The second phase consisted of structured focus group discussions conducted in the form of on-line workshop in February 2025 with experts from companies including Google, Cognizant and EPAM, aimed at evaluating and refining the proposed framework. These discussions enabled participants to engage critically with the initial findings and contribute insights on model clarity, relevance, and potential applications across different domains. The third phase focused on gathering expert perspectives through 20 semi-structured interviews conducted between February and April 2025. Participants included AI researchers, developers, sustainability officers, and industry leaders invited by dr2.ai research team. The interviews explored current applications and future opportunities of AI within personal, professional, and urban domains, described in the framework as lenses. Additionally, participants reflected on key challenges, particularly those related to environmental sustainability and data privacy.

The participants from both focus group discussions and semi-structured interviews consisted of specialists who had at least three years of professional experience in ML or AI. Majority of them had in the portfolio projects concerning wearables and other consumer products including smartphones and smartwatches as well as in-car communication systems and smart home technologies. The participants worked on projects that required collaboration between teams from Europe, North America and the Asia-Pacific region. They demonstrated both worldwide perspectives and understanding of worldwide trends and AI-based product and service interconnectivity. The study's technical nature and exploratory approach did not require gender-based selection criteria because the research focused on professional qualifications and topic relevance. A purposive sampling strategy was applied. The dr2.ai research team used professional networks including LinkedIn platform, academic collaborations between Warsaw School of Economics and other universities, and industry partnerships to identify participants for the study. In parallel with these research phases, a broad range of products, services, and projects from leading technology companies were analysed. They were assessed based on criteria including technological maturity and environmental impact.

Discussed study faced constraints, including limited access to proprietary specifications, rapid technology evolution during the study period, varying levels of public information availability. Triangulation was achieved by cross-referencing information from the literature review with workshop findings and results of interviews. The methodology was designed to provide a comprehensive understanding of the current state of the art in advanced technologies while identifying patterns in the evolution toward agentic environments, with a particular focus on efficiency and potential for cross-domain integration.

## 3. The Evolution of Technological Environments

Technology has undergone remarkable progress in the past decades, significantly influencing how it is utilized in human environments [15]. This evolution, from traditional IT to intelligent, human-centric environments, was enabled by advances in AI, sensors, and human-computer interaction [16]. Recent approaches focus on empowering human abilities and integrating seamlessly with the physical world [17]. Nevertheless, the emergence of GenAI and compound AI systems has the potential to revolutionize, in the short term, human capability to create environments that are cognitive, adaptable, and, foremost, sustainable [18]. This approach corresponds with Sustainable Development Goal 11 (SDG 11) which is focused on cities as resilient, inclusive and ultimately sustainable urban spaces [19]. This is important, especially in the context of demographic shift and use of resources. Already over half of the world's population lives in cites and forecast show a growing trend towards 70% by 2050 [20]. It means a growing pressure on urban infrastructure, use of energy and water as well as transportation and numerous other services. Therefore, solutions based on AI focused on energy efficiency and circular principles could bring benefits, especially alignment with SDG 11 sub-targets (11.2 mobility, 11.6 air quality & waste, 11.b resilience planning) [19], which provides quantifiable assessment criteria for future evaluation.

Smart environments emerged with the rise of smart buildings and cities in the late 20th century [20]. As Cook and Das [21] noted, the goal of smart environments is to create a space that brings computation into the physical world, allowing it to become part of the normal activities and tasks of everyday life. They are built on interconnected digital systems, including IoT devices and sensors, to enhance efficiency in various areas, from smart buildings, traffic management, and lighting to entire districts, such as Songdo IBD, or even cities like Abu Dhabi's smart city [22]. As a result, smart environments are seen by many cities as integral to efforts to create sustainable living and working spaces, aligned with broader ecological goals and even sustainable development [23].

Another stage of transformation going beyond smart environments involves ambient intelligence (AmI). It was developed on the notion that the integration of technology into everyday life should be seamless and intuitive [24]. These systems anticipate user needs without explicit commands. As Weber and Aarts [25] described, Ambient intelligence refers to electronic environments that are sensitive and responsive to the presence of people. The advances in AmI have been driven by sensor technologies, machine learning, and context-aware computing. These technologies allow dynamic adaptation to change in users' behavioural patterns and offer context-aware services that enhance comfort and safety, for instance, Amazon's Alexa, Google Assistant, and Apple's Siri.

Intelligent environments are spaces with embedded systems and information and communication technologies creating interactive spaces that bring computation into the physical world and enhance occupants' experiences. Intelligent environments can integrate multiple functionalities seamlessly [26]. They employ cognitive modelling, adaptive learning, and multimodal interaction to enhance user experience [27]. Intelligent environments represent the convergence of AI and IoT. Burzagli [28] further this integration by stating that intelligent environments are formed by the convergence of several computing areas: ubiquitous and pervasive computing, ambient intelligence, context awareness, sensor networks, artificial intelligence, and human-computer interaction. In the context of sustainability, they have the potential to contribute to the optimization of resource utilisation. Highlighting this potential, Bibri and Krogstie [29] note, the integration of big data analytics and context-aware computing into intelligent environments offers tremendous opportunities for advancing environmental sustainability in cities through optimized energy consumption, improved waste management, and enhanced urban ecosystem services.

All the concepts discussed so far appear greatly limited when compared to recent innovations, including Generative AI. One of their limitations is the lack of a holistic and cross-domain approach. Also, they are not yet equipped with the advanced reasoning, learning and problem-solving capabilities that Generative AI promises [30]. These solutions lack the autonomy to operate without extensive human input or oversight, particularly in complex scenarios requiring nuanced decision-making across multiple domains [31]. For a comprehensive comparison of environments' evolution, please refer to Table 1, which presents the main characteristics, environmental impact, key technologies used, and possible applications.

**Table 1.** Comparative analysis of smart, ambient, intelligent and agentic environment paradigms.

|  | Smart Environments | Ambient Intelligence | Intelligent Environments | Agentic Environments |
|---|---|---|---|---|
| **Definition** | Environments enhanced by interconnected digital systems to improve efficiency, comfort and decision-making; utilize sensors, actuators and data analytics to monitor and respond to various conditions. | Environments that are sensitive, adaptive and responsive to the presence of people; aim to create a pervasive and intuitive technological presence. | Environments that leverage cognitive processes and advanced technologies to create dynamic and responsive spaces that adapt to and support occupants using context. | Physical and digital spaces incorporating AI agents, powered by GenAI models, creating responsive, sustainable and intelligent ecosystems that actively engage with humans and the surroundings to provide comprehensive assistance. |
| **Functionality** | Automation of tasks, real-time monitoring and data collection for improved decision-making. For instance, energy management, predictive maintenance. | Context-aware services that adapt to user behaviour and preferences, often without direct user input. For instance, seamless device integration. | Dynamic interactions that anticipate user desires, integrating multiple functionalities seamlessly. For instance, cognitive modelling, adaptive learning. | Proactive assistance, contextual intelligence, seamless integration of multiple agents working collaboratively across different domains. For instance, cross-domain optimization, predictive personalization. |
| **Sustainability impact** | Focus on energy optimization through smart metering and adaptive controls; resource-efficient operations based on occupancy and usage patterns; integration with renewable energy systems; smart waste management systems. | Energy efficiency through automated context-aware control systems; improved indoor environmental quality. Promotion of sustainable behaviours. Health and well-being: through ambient monitoring. | AI-driven predictive resource management; support for circular economy practices; enhanced accessibility and adaptability to diverse needs. Ecosystem Integration: Improved urban planning and management. | Holistic sustainability management across personal, professional and urban context; AI-driven balancing of human needs with environmental impact; promotion of circular economy practices through agent-mediated exchanges. |
| **User Interaction** | Primarily through apps, voice commands, touch interfaces, smart devices and | Minimal explicit interaction: system responds to user context and implicit needs; gesture | Natural interfaces, gesture recognition, voice commands, brain-computer | Natural language interaction, proactive assistance, multi-modal engagement (voice, text, |
|  | dashboards for data visualization. | recognition, ambient displays. | interfaces, augmented and mixed reality. | gesture), highly personalized interfaces. |
| Key Technologies | IoT devices, sensors, actuators, cloud computing, big data analytics, wireless communication protocols. | Sensor fusion, machine learning, tinyML, context-aware computing, embedded sensors, edge computing, low-power communication. | AI, machine learning, cognitive computing, advanced sensors, adaptive systems, natural interfaces, sensor networks. | Large Language Models, multi-modal interfaces and models, multi-agent systems, natural language processing and understanding (NLP, NLU), edge computing, federated learning. |
| Applications | Environmental monitoring, Smart buildings – HVAC management Smart cities – traffic and waste management. | Smart Assistants Smart Homes - Environments that adjust settings based on user presence and preferences. Healthcare monitoring | Cognitive Smart Homes - Environments for optimal comfort and efficiency. Interactive workspaces - adapt to users' needs. | Personalized AI assistants managing daily tasks; collaborative workplace agents facilitating complex projects; urban-scale coordination for city services and infrastructure. |

Source: Authors' own elaboration based on [18,21,24–26,28,30,32].

### 4. Transformative Enablers: Generative AI, AI Agents and Edge intelligence

Early generative models like generative adversarial networks (GANs) and variational autoencoders (VAEs) showed potential for creating original visual and audio content, while they create original content by learning from data [33]. Language models used currently evolved from sequence to sequence models and recurrent neural networks (RNNs) [34] to transformer architecture [35], enabling development of large language models (LLMs). In the text and image domains, the field is strongly led nowadays by evolutions of the GPT model from OpenAI [36], significantly enhancing its ability to reason and generate coherent output. Other widely used LLM models, including Llama [37], Gemini [38], and Claude [39], have further augmented this trend. Models like Whisper [40] and AudioCraft [41] drive realistic text-to-speech and music generation. In the video domain, SORA has emerged as a groundbreaking model that offers the ability to develop dynamic visual stories by intelligently modelling motion and content, thus expanding the horizons of video generation [42]. LLMs reflect the link between language and human cognition [43]. Enhancing these models with additional capabilities, including sensor data, memory integration, and collaborative functionalities, will lead to more dynamic innovations bridging human-computer interactions [44].

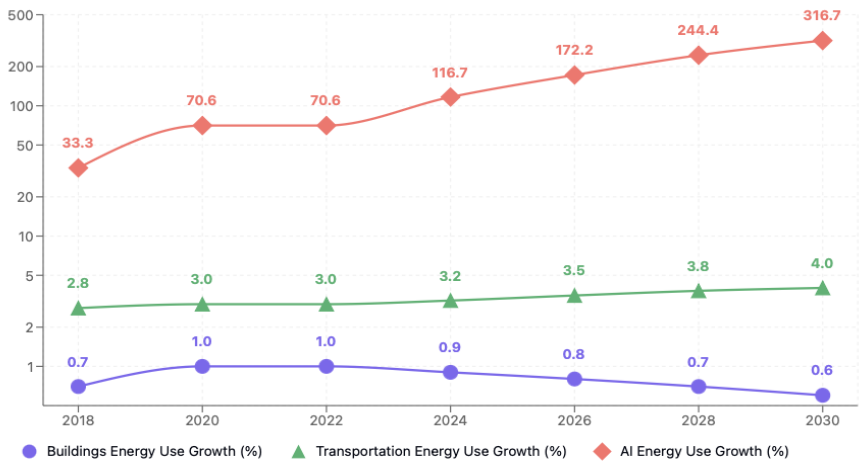

**Figure 1.** Yearly Increase (%) in Energy Consumption Across Sectors (2018-2030).
Source: Authors' own elaboration based on IEA [7].

Although one can notice the progress in foundation models like GPT, Claude, and Llama, key technical and ethical issues remain present. This includes hallucinations, inherent biases, and concerns around training data and output ownership [42,45,46]. Their training and inference require significant computational power, leading to major energy and environmental costs. Training GPT-4 (1.8T parameters) requires approximately 2,000 Nvidia Blackwell GPUs, consuming around 4 MWh, equivalent to the annual energy use of about 3,200 U.S. homes [36,47,48]. As presented in figure 1, energy use by AI is expected to grow significantly year over year [7], outpacing relatively stable trends in buildings and transportation [49].

Model scale continues to rise, Llama 3.1 (405B parameters) is a massive leap from Llama 2's 70B in just one year. GPT-4 exemplifies this with 16 models, each with approximately 111B parameters, ten times more than GPT-3's 175B. The rise in LLM size is driven by goals like better performance, enhanced multimodality, generalization, and progress toward AGI [50]. This type of rapid growth requires advanced training techniques and rigorous data curation [51,52]. In contrast, small language models (SLMs), under 8B parameters, are emerging as a sustainable alternative [53]. SLMs are modular, decentralized, and run locally on phones or laptops. This reduces cloud GPU emissions and improves AI accessibility [54,55]. A key development enabling decentralized use of LLMs is the AI Agent framework [56]. It allows AI to reason beyond pre-training by connecting with environments and other agents. These systems are both perceptive and responsive, adapting to user needs and context. Core components include the agent, planning mechanisms, memory, and specialized tools [57]. Figure 2 illustrates a proposed agent architecture for agentic environments. This architecture is evolving toward multi-agent systems [58], where multiple domain-specific LLM-powered agents collaborate on complex tasks. These systems incorporate belief representation, task assignment, communication, and joint decision-making. Components include a world environment and a communication protocol, which enable agents' interactions and coordination. Unlike monolithic models, LLM agent systems reflect a compound AI paradigm, modular, decentralized, and capable of reasoning-action loops, either autonomously or with humans in the loop [59].

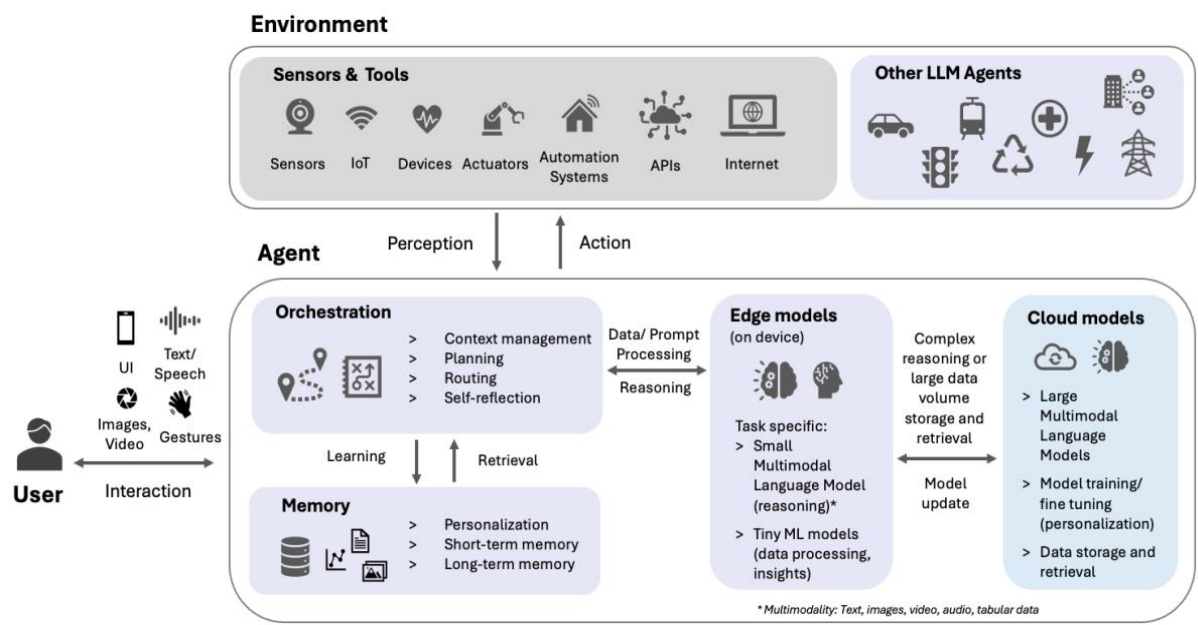

**Figure 2.** LLM multi-agent architecture for agentic environments.
Source: Authors' own elaboration based on [56,60].

The deployment of compound AI systems requires edge-driven or hybrid edge-cloud models to achieve personalization and privacy and sustainability goals [61]. Edge intelligence functions through efficient real-time secure processing which enables AI model inference directly on devices including smartphones and wearables and IoT systems [62]. The integration of LLMs at the edge not only transforms the landscape for innovation but becomes critical for applications that require low latency and privacy, such as personal assistants, smart homes, automotive technologies, and healthcare [63]. These edge-based

AI agents can deliver human-like language interactions and power context-aware autonomous interactions, all while providing deep personalization through on-device fine-tuning, optimisation and long-term memory retention [64]. The increasing computational power edge hardware with decreasing its operational costs may further drive democratization and deployment of generative AI on devices like smartphones [65]. A key challenge in deploying LLMs on edge devices lies in their size and computational demands. Models like GPT-4 and LLaMA-3, with billions or even trillions of parameters, are too resource-intensive for edge environments due to their GPU requirements. Even scaled-down models, such as Microsoft's Phi3-mini with 3.8 billion parameters, still require high-end hardware, making them impractical for most edge devices [66]. One approach is to use smaller, specialised solutions like Apple's AFM series [67], optimised for edge hardware, though these models are limited in handling open-ended tasks and rely on predefined workflows. Model compression techniques including quantization reduce both model size and computational requirements but result in decreased accuracy and sensitivity levels [68]. The recent advancements in in small and sparsely activated language models (SLMs/SALMs) and neuromorphic edge hardware, like Groq Language Processing Unit, show that it is possible to conduct tasks more efficiently from an energy perspective compared to GPUs. Integration of these findings with agentic environments could open a path to net energy savings and sustainable agentic solutions.

## 5. Towards Agentic Environments: The Conceptual Framework

Based on an extensive literature review, expert workshops, and semi-structured interviews conducted with AI specialists, urban planners, and technologists, including a focus group held in February 2025, 8 experts led by dr2.ai research team and Warsaw School of Economics academics worked on a conceptual framework of agentic environments. This framework reflects the insights from theory, industry practice, and stakeholder expectations. Authors propose a concept of agentic environments that introduces a novel and advanced stage in the evolution of intelligent spaces, focusing on the sustainable utilization of recent advancements in AI and the integration of cross-domain human-centric environments. Agentic environments infuse physical and digital spaces, offering ecosystems that not only respond to human interaction but can understand the surrounding context, anticipate needs, and engage with both individual and global environments, promoting enhanced quality of life and balanced sustainability. From the semi-structured interviews, the proposed framework was further informed by these insights. Many experts stressed the role of multi-agent systems in developing adaptable and resource-efficient solutions across personal, professional, and urban domains, referred to as 'lenses' within the framework.

In this context, the novelty of agentic environments lies in their full utilization of compound AI architectures, a multi-agent framework, and a combination of edge-cloud multimodal language models that deliver complex AI reasoning. This enables the creation of dynamic ecosystems where AI agents actively manage and optimize life and infrastructure, from personal living spaces to large-scale urban systems. Agentic environments implement sustainability strategies by interpreting explicit and implicit human signals, enabling personalized, seamless interaction and optimization, while supporting circular economy practices responsive to both global and local data. By shifting inference and model learning to more efficient, privacy-oriented edge agents, agentic environments reduce reliance on data centres and promote sustainable AI. At the core of this system are AI agents that fully leverage LLMs' capabilities in multimodal perception, contextual reasoning, and environment interaction. The architecture is modular, versatile, and designed for seamless deployment across diverse environments, from personal assistants to smart cities.

The agents can work together through cooperative, hierarchical or mixed strategies to solve domain and task problems autonomously or in teams based on context. In a cooperative model, agents share domain knowledge to solve complex problems, while hierarchical models delegate tasks through a layered structure. AI agents can be tailored with sensors, memory, tool integrations and user-specific learning mechanisms. They can operate on edge or hybrid systems, leveraging various LLMs depending on context, modality, and processing needs. Deep customisation is driven by user input, contextual interaction, and federated fine-tuning, allowing fast adaptation and increased effectiveness. By using edge computing local processing, reducing latency and decentralising computation are possible. This leads to improved responsiveness and resilience. For instance, smart homes can use local LLMs for energy efficiency, while more demanding tasks like traffic management benefit from cloud-based reasoning. Multimodality allows LLMs to process diverse inputs including text, voice, images, sensor data, at once. According to experts in the focus group, this combination of modularity, edge-driven architecture, distributed intelligence, and deep personalization supports scalable, sustainable solutions. It accommodates growing computational needs across contexts including smart homes, workplaces, or cities, while maintaining strong privacy and data protection.

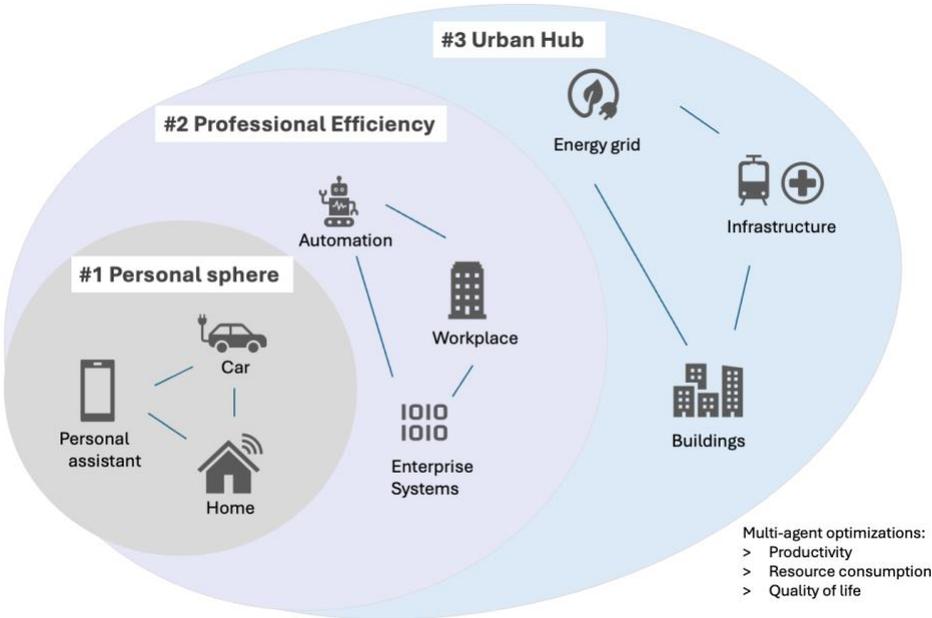

**Figure 3.** The sustainable agentic environment lenses. LLM multi-agents integration across personal, professional and urban sphere.
Source: Authors' own elaboration.

To illustrate the concept, the authors focus on three spheres, referred to as agentic lenses: beginning with private settings, then moving to work and public settings, and concluding with an overarching city setting. Please refer to Figure 3 for a graphic representation of these lenses and their interconnectivity. Figure 4 describes how the lenses and cross-domain agents collaborate to enable the agentic environments.

**Lens 1: Personal Sphere - Human-Centred Applications**
First lens, personal sphere, is a fundamental component of the proposed architecture. It includes a range of human-centred applications such as personal assistants, smart homes and smart vehicles. Currently, solutions in the personal sphere cover range spectrum of use cases, including voice-activated digital assistants like Amazon Alexa or Google Assistant, Nest thermostats or the Philips smart home automation solution. They facilitate tasks like managing schedules or controlling smart home systems. Current trends involve conversational models like GPT in Apple Intelligence, enabling Siri to automate smart home controls and manage text with on-device privacy. Also, Meta's and Even Realities GenAI-powered smart glasses [69], and Qualcomm's generative AI in V2X communication [70] expand use cases. Smart glasses use LLMs to interpret visual data; V2X systems offer real-time recommendations to drivers [71]. Despite evolution, solutions function mostly in silos due to technological, legal and behavioural reasons. This hinders deep contextual understanding. An agent-based approach may help, using AI agents to manage environments like phone, PC and smartwatch. For example, an assistant can create recommendations for smart systems in home and car, aiding decisions like ordering groceries or commuting. Wearable health devices can communicate with home or vehicle systems. For example, if stress is detected, a smartwatch may suggest rest, reroute the car or adjust lighting/music at home. The home system can inform the vehicle about parking or charging stations. As remote work merges work and home, AI agents become crucial. During work tasks, the system can coordinate home activities like grocery ordering or managing heating and lighting.

**Lens 2: Professional Efficiency**

The next lens presents the environment in which AI agents are applied in professional and commercial settings to boost productivity, streamline operations, and enhance security. From workplace to healthcare and emergency responses, AI and AI agents are used as knowledge and decision support tools.

Current solutions for productivity and automation include Microsoft's Copilot for Office, GitHub Copilot for coding, ChatGPT for content creation, creative tasks and research [72]. These tools are embedded in systems or deployed as stand-alone tools. For knowledge management, Perplexity is used for research and retrieval [73]. Atlassian's Rovo integrates data, offers corporate search, and helps manage tasks [74]. GenAI also supports customer service, Zendesk uses AI for ticket triage, prioritization and responses [75]. RPA is evolving too: UiPath enables interpretation of interfaces and ML in workflows; Automation Anywhere uses AI bots to process unstructured data and language [76]. Despite progress, many solutions operate within limited scopes and rely on pre-trained models without real-time learning. Solutions work in isolation. Integration between workplace (Lens 2) and personal sphere (Lens 1) offers potential, e.g. calendar syncing, commute or meal planning. Synergy of health and well-being data also opens applications for agents. Integration and communication with external devices must occur with user consent, or in critical health situations. Smartwatches (Lens 1) gather personalized data like steps, heart rate, temperature, glucose. These wearables are often with users and useful in critical situations. Personal agents could analyse this data and contact emergency services (Lens 2) or guide users to improve well-being (Lens 1). Healthcare data can monitor health at work, detecting stress or fatigue. Agents could then suggest stress relief or adjust workload for mental and physical health.

**Lens 3: Urban Hub – Infrastructure, Operations and Environmental Systems**

The last lens is focused on larger-scale projects and infrastructure, including smart buildings, transportation systems, and smart cities. Key areas include building management systems (BMS), urban transportation, public safety, and utilities. In intelligent buildings, companies like Honeywell implement AI for property and energy management. Platforms like Forge enable monitoring of operations [77]. Siemens' Desigo CC streamlines lighting, HVAC, elevators, while Johnson Controls' Metasys BMS enhances security with real-time monitoring [78]. In smart cities, AI is used for traffic, energy grids, waste, environmental and safety monitoring. Google's Green Light optimizes traffic in 13 cities, reducing emissions [79]. Buenos Aires uses 'Boti', an AI chatbot, for citizen engagement [80]. Waymo's AVs in San Francisco and Phoenix prevent CO2 emissions [81], though concerns exist about car dependency.

On energy, ABB and Copenhagen collaborate on carbon-neutral living with AI-managed renewable energy [82]. In resilience, Lisbon uses Bentley Systems' AI-driven flood simulator for climate adaptation [83]. NASA and IBM developed 'Prithvi-weather-climate', an AI model trained on 40 years of data, for severe weather forecasting [84]. The systems operate similarly to other lenses because they do not integrate complex reasoning for broad optimization. The main obstacles include interoperability issues together with data privacy concerns and scalability problems. The development of integrated frameworks becomes essential to support collaboration between urban areas and professional and personal domains. GenAI-driven agentic environments can integrate data from transportation and energy and safety domains to support complex reasoning and complete understanding.

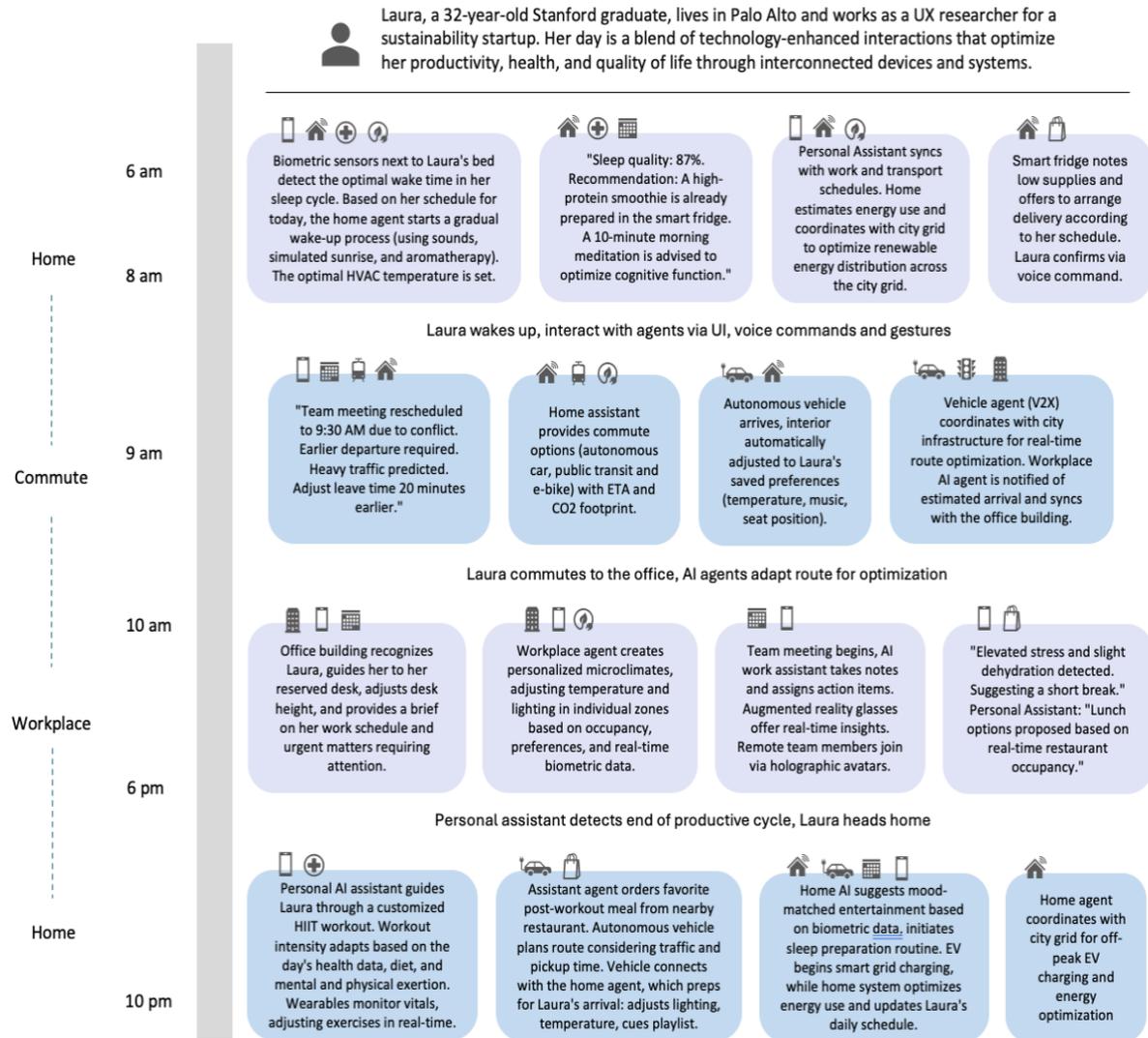

**Figure 4.** Daily routine scenario presenting cross-lens agentic environment interactions.
Source: Authors' own elaboration based on the literature review and focus group results and feedback from semi-structured interviews.

## 6. Discussion and future prospects

Discussion on the models and futures of agentic environments was conducted as part of the research methodology, the developed model was discussed with two dozen of experts during semi-structured interviews. The general view was that the agentic environments framework that integrates GenAI agents across personal, workplace, and urban domains, enabling the sustainable use of technological advancements, holds significant potential for enhancing broadly understood efficiency and sustainability at both local and coordinated urban levels, nonetheless, the proposed solution carries certain inherent challenges. Among the most frequently mentioned were concerns related to energy consumption, data privacy, and long-term sustainability of agent-based AI systems, experts pointed out that although the proposed architecture focuses on edge computing and decentralization, many current solutions still rely on high-computational cloud infrastructure, which is energy-intensive and environmentally costly, several interviewees indicated that energy demand may further rise with the increasing use of LLMs in everyday scenarios.

One of the key critical reflections shared by experts is what they referred to as the sustainability paradox, where more efficient systems and services may actually increase the total resource consumption due to higher demand and wider deployment [85], for instance, the Songdo smart city project in South Korea, designed as a green, high-tech city, has recorded significantly higher than expected energy use due to its dense sensor networks and AI infrastructure, Barcelona's smart traffic systems, while reducing congestion, have reportedly encouraged increased vehicle usage, similarly, Singapore's predictive AI infrastructure, designed to streamline operations, led to increased energy use from data storage and computing. From expert interviews, several suggestions emerged, to mitigate environmental costs, many advocated the adoption of low-power AI hardware, more efficient model training, and federated learning to reduce central computation, they highlighted the importance of developing foundational models designed for edge deployment, which would drastically lower dependency on cloud GPUs and reduce emissions from data centres.

In addition to energy concerns, data privacy and security were considered critical, as AI agents increasingly enter personal homes and urban infrastructures, the data they process becomes more sensitive, multiple experts raised alarms about existing vulnerabilities, many already identified in commercially available devices, during interviews, several examples were cited, smart home hubs such as Amazon Echo or Google Nest, as well as assistants like Apple Siri, have faced investigations over data misuse and potential for unauthorized access. Another point raised by participants was that users tend to rely on cloud-based AI systems even for basic tasks, like writing an email or creating to-do lists, many such activities could be supported by local, less resource-intensive edge agents, but due to ease-of-use, users default to ChatGPT or Perplexity, which indirectly increases environmental costs. The environmental impact of model training and inference was highlighted as well, some interviewees cited studies [86] showing that training one large language model can result in $CO_2$ emissions comparable to five cars' lifetime emissions, therefore, experts proposed policy-level regulations, including carbon accounting for AI models, and stricter guidelines for environmental audits of AI systems. The future direction should focus on transitioning to renewable energy sources for powering data centres and AI infrastructures, according to experts, a combination of wind, solar, and nuclear energy can reduce carbon emissions, promising technologies mentioned included Groq's low-energy inference chips and IBM's neuromorphic processors (like Loihi), which may facilitate sustainable on-device computing.

Future research needs to focus on developing cross-lens and cross-domain agentic systems especially in urban environments which are complex, data-rich and politically sensitive. Achieving effective integration will require standardised frameworks, robust privacy and security protocols, and sustainable business models to support the necessary infrastructure. The development of Physical AI and World Models and Foundation Models that process multimodal sensor data and Agentic Digital Twins should be prioritised to enhance intelligent systems' ability to perceive and interpret and interact with dynamic real-world contexts.

## 7. Conclusions

The evolution of technological environments discussed in the article reflects a shift towards artificial intelligence. Starting from traditional IT systems to intelligent, adaptive, and human-centred solutions, authors focused on agentic environments, which address important limitations of traditional intelligent systems, including growing energy demand, reliance on centralised cloud infrastructure, and data security risks. Their proactive and adaptive nature, as well real-time contextual learning, can contribute to optimisation of resources, as well as improved user experience, and reduced environmental impact. In contrary to current siloed systems, agentic environments integrate across domains using modular AI agents and edge computing. They also offer deeper personalization and lower energy consumption.

Experts interviewed as part of the research emphasized both the potential and challenges of agentic environments. The important concern was the sustainability paradox, where increased efficiency can lead to greater overall resource consumption. Examples from Songdo, Singapore, and Barcelona illustrate that even smart city solutions may increase energy use due to infrastructure and usage scale. Privacy and security were also flagged as critical issues, particularly as AI agents become embedded in homes and public systems. In this context, the proposed framework provides a roadmap for responsible AI integration, in which focus shifts from purely data-driven efficiency to a more holistic model encompassing human needs, environmental constraints, and ethical concerns. From mobile assistants to city wide systems, agents facilitate intelligent actions grounded in sustainable practice.

Authors of the study recommend further investigation into the development of low impact AI architectures, hybrid edge-cloud systems, and proof of concept deployments across the three defined lenses. Industry should consider adoption of transparent carbon reporting and possibly introduce "green AI" certifications like other industries already did for the products, including property industry and green buildings corticates. Therefore, energy efficient AI models, and AI-specific sustainability metrics will be essential for further development of sustainable advanced technologies.

**Acknowledgements**

The authors would like to thank all the experts who generously contributed their time and insights during the focus group discussions and interviews. We are especially grateful to the AI researchers and professionals surveyed or interviewed for their valuable input, which significantly enriched the research.